\documentclass[fleqn,usenatbib]{mnras}

\usepackage{newtxtext,newtxmath}
\usepackage[T1]{fontenc}

\DeclareRobustCommand{\VAN}[3]{#2}
\let\VANthebibliography\thebibliography
\def\thebibliography{\DeclareRobustCommand{\VAN}[3]{##3}\VANthebibliography}

\usepackage{graphicx}	% Including figure files
\usepackage{amsmath}	% Advanced maths commands
\newcommand{\Ni}{\ensuremath{^{56}\mathrm{Ni}}}
\newcommand{\Co}{\ensuremath{^{56}\mathrm{Co}}}

\newcommand{\Msun}{\ensuremath{\mathrm{M}_\odot}}
\newcommand{\Rsun}{\ensuremath{\mathrm{R}_\odot}}

\graphicspath{{./}{figures/}}

\title[GRSN properties]{
Observational properties of a general relativistic instability supernova from a primordial supermassive star
}

\author[T. J. Moriya et al.]{
Takashi J. Moriya,$^{1,2}$\thanks{E-mail: takashi.moriya@nao.ac.jp (TJM)}
Ke-Jung Chen,$^{3}$
Kimihiko Nakajima,$^{1}$
Nozomu Tominaga,$^{4,5}$ and \newauthor
Sergei I. Blinnikov$^{6,7,5}$
\\
$^{1}$National Astronomical Observatory of Japan, National Institutes of Natural Sciences, 2-21-1 Osawa, Mitaka, Tokyo 181-8588, Japan \\
$^{2}$School of Physics and Astronomy, Faculty of Science, Monash University, Clayton, Victoria 3800, Australia \\
$^{3}$ Institute of Astronomy and Astrophysics, Academia Sinica, Taipei 10617, Taiwan \\
$^{4}$ Department of Physics, Faculty of Science and Engineering, Konan University, 8-9-1 Okamoto, Kobe, Hyogo 658-8501, Japan \\
$^{5}$ Kavli Institute for the Physics and Mathematics of the Universe (WPI), The University of Tokyo Institutes for Advanced Study, The University of Tokyo, \\ 5-1-5 Kashiwanoha, Kashiwa, Chiba 277-8583, Japan \\
$^{6}$ National Research Center "Kurchatov institute", Institute for Theoretical and Experimental Physics (ITEP), 117218 Moscow, Russia \\
$^{7}$ Dukhov Automatics Research Institute (VNIIA), Suschevskaya 22, 127055 Moscow, Russia
}

\date{Accepted 2021 February 26. Received 2021 February 26; in original form 2020 June 22}

\pubyear{2021}

\begin{document}
\label{firstpage}
\pagerange{\pageref{firstpage}--\pageref{lastpage}}
\maketitle

\begin{abstract}
We present the expected observational properties of a general relativistic instability supernova (GRSN) from the 55,500~\Msun\ primordial (Population~III) star. Supermassive stars exceeding $10^4~\Msun$ may exist in the early Universe. They are generally considered to collapse through the general relativistic instability to be seed black holes to form supermassive ($\sim 10^9~\Msun$) black holes observed as high-redshift quasars. Some of them, however, may explode as GRSNe if the explosive helium burning unbinds the supermassive stars following the collapse triggered by the general relativistic instability. We perform the radiation hydrodynamics simulation of the GRSN starting shortly before the shock breakout. We find that the GRSN is characterized by a long-lasting (550~d) luminous ($1.5\times 10^{44}~\mathrm{erg~s^{-1}}$) plateau phase with the photospheric temperature of around 5000~K in the rest frame. The plateau phase lasts for decades when it appears at high redshifts and it will likely be observed as a persistent source in the future deep near-infrared imaging surveys. Especially, the near-infrared images reaching 29 AB~magnitude that can be obtained by Galaxy and Reionization EXplorer (G-REX) and James Webb Space Telescope (JWST) allow us to identify GRSNe up to $z\simeq 15$. Deeper images enable us to discover GRSNe at even higher redshifts. Having extremely red color, they can be distinguished from other persistent sources such as high-redshift galaxies by using color information. 
%Finally, the cooling phase of the GRSN shortly after the shock breakout can be observed as a transient even if they appear at $z \gtrsim 15$. The rise time at this phase is about a month in the observer frame and the peak near-infrared magnitude is around $28-28.5~\mathrm{AB~magnitude}$. 
We conclude that the deep near-infrared images are able to constrain the existence of GRSNe from the primordial supermassive stars in the Universe even without the time domain information.
\end{abstract}

\begin{keywords}
supernovae: general -- dark ages, reionization, first stars -- early Universe -- quasars: supermassive black holes -- stars: Population III
\end{keywords}

%%%%%%%%%%%%%%%%%%%%%%%%%%%%%%%%%%%%%%%%%%%%%%%%%%

%%%%%%%%%%%%%%%%% BODY OF PAPER %%%%%%%%%%%%%%%%%%

\section{Introduction}
Supermassive black holes (SMBHs) exceeding $\sim 10^9~\Msun$ are known to exist at $z> 6$ through high-redshift quasar observations \citep{fan2001qso,fan2003qso,willott2007qso,mortlock2011qso,morganson2012qso,kashikawa2015qso,wu2015qso,banados2016qso,banados2018qso,matsuoka2019qso}. The age of the Universe at $z > 6$ is less than 1~Gyr. Forming SMBHs in 1~Gyr is very challenging and it is one of the frontiers in the modern astrophysics (see \citealt{inayoshi2020smbh,2020SSRv..216...48H} for recent reviews). 

One possible path to form SMBHs in such a short timescale is through Population~III supermassive stars (SMSs) having $10^4-10^6~\Msun$. Although the typical mass of Population~III stars is predicted to be much less than $10^4-10^6~\Msun$ \citep[e.g.,][]{mckee2008popiiirf,hosokawa2011popiiifb,susa2014popiiiimf,hirano2015popiiiimf,sugimura2020}, SMSs can be formed by preventing $\mathrm{H_2}$ cooling through, e.g., intense the Lyman-Werner ultraviolet background radiation photodissociating $\mathrm{H_2}$ \citep[e.g.,][]{omukai2001,oh2002,bromm2003,sugimura2014,chon2016,chon2018,regan2017}. SMSs are predicted to collapse through general relativistic instability \citep[e.g.,][]{iben1963,chandrasekhar1964,fowler1966,osaki1966,shibata2002,shibata2016,umeda2016,uchida2017,nagele2020}, forming seed massive black holes (BHs). SMSs can be formed at $z\simeq 15-20$ \citep[e.g.,][]{agarwal2012,dijkstra2014,habouzit2016} to leave the seed BHs and they can grow to SMBHs in 1~Gyr to explain the existence of SMBHs at $z> 6$.

The general relativistic instability of SMSs does not necessarily lead to their collapse to BHs \citep{fuller1986,montero2012,chen2014gre,nagele2020}. In particular, \citet{chen2014gre} showed that a non-rotating Population~III SMS with 55,500~\Msun\ explodes as a supernova (SN) through the explosive helium burning following the collapse. Although SMSs causing such a general relativistic instability SNe (GRSNe) may be limited in a small mass range, they could be bright enough to be observed in the future near-infrared (NIR) transient surveys \citep{whalen2013grsn}. They also have distinctive chemical signatures that can be traced by the stellar archaeology of unusual extreme metal poor stars \citep{johnson2013}.

In this paper, we investigate the observational properties of the GRSN from the 55,500~\Msun\ SMS presented by \citet{chen2014gre}. The observational properties of the GRSN were previously investigated by \citet{whalen2013grsn}. They showed that future NIR transient surveys can discover the GRSN. In this work, we present our new radiation hydrodynamics simulation of the GRSN and suggest that it is not required to conduct transient surveys to discover the GRSN given its extremely long timescale keeping its brightness. Having deep multi-band images in NIR would be enough to identify them. 

The rest of the paper is organized as follows. We first show our method of the GRSN investigation in Section~\ref{sec:method}. Then we discuss its properties in the rest frame in Section~\ref{sec:restframeproperties}. Then we show its observational properties when it appears in the early Universe and discuss how to find it in Section~\ref{sec:howtodiscover}. We conclude this paper in Section~\ref{sec:summary}. We adopt the standard $\Lambda$CDM cosmology with $H_0=70~\mathrm{km~s^{-1}}~\mathrm{Mpc^{-1}}$, $\Omega_M = 0.3$, and $\Omega_\Lambda = 0.7$ throughout this paper.

\section{Method}\label{sec:method}
\subsection{GRSN model}
We take the GRSN explosion model of the 55,500~\Msun\ Population~III star presented by \citet{chen2014gre}. We adopt their two-dimensional explosion model. The evolution of the 55,500~\Msun\ star is followed by using the \texttt{KEPLER} one-dimensional stellar evolution code \citep{weaver1978kepler}. The star is found to explode with the one-dimensional calculation with \texttt{KEPLER} \citep{chen2014gre}. The radius of the star at the time of the explosion is 256~\Rsun.

The one-dimensional stellar structure at 600~s before the maximum compression at the center in the one-dimensional calculation, which is sufficiently long before the explosive burning leading to the explosion, is transferred to the multi-dimensional hydrodynamics code \texttt{CASTRO} \citep{almgren2010castro,zhang2011castro}. The
subsequent hydrodynamic evolution with nucleosynthesis is followed until shortly before the shock breakout by two-dimensional hydrodynamics.
We adopt the result from the multi-dimensional simulation because of the strong mixing found in the GRSN which affect the abundance profile and, therefore, synthetic LCs \citep{chen2014gre}.
The explosion energy is $9\times 10^{54}~\mathrm{erg}$. 
Because the core temperature ($<10^9$~K) and density ($<10^3~\mathrm{g~cm^{-3}}$) is low at the explosive helium burning, 
little \Ni\ is produced during the explosion and the total \Ni\ mass in the ejecta is less than 0.1~\Msun.

\begin{figure}
	\includegraphics[width=\columnwidth]{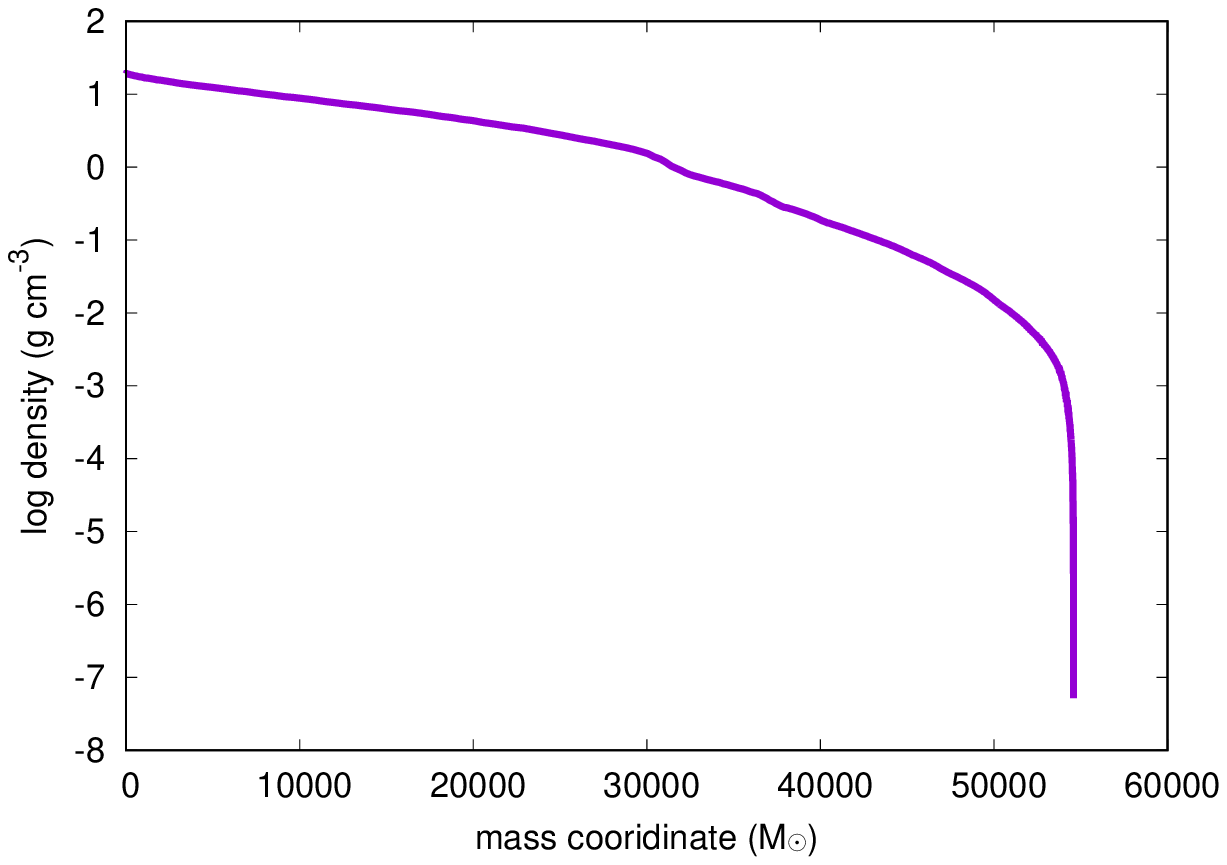}
	\includegraphics[width=\columnwidth]{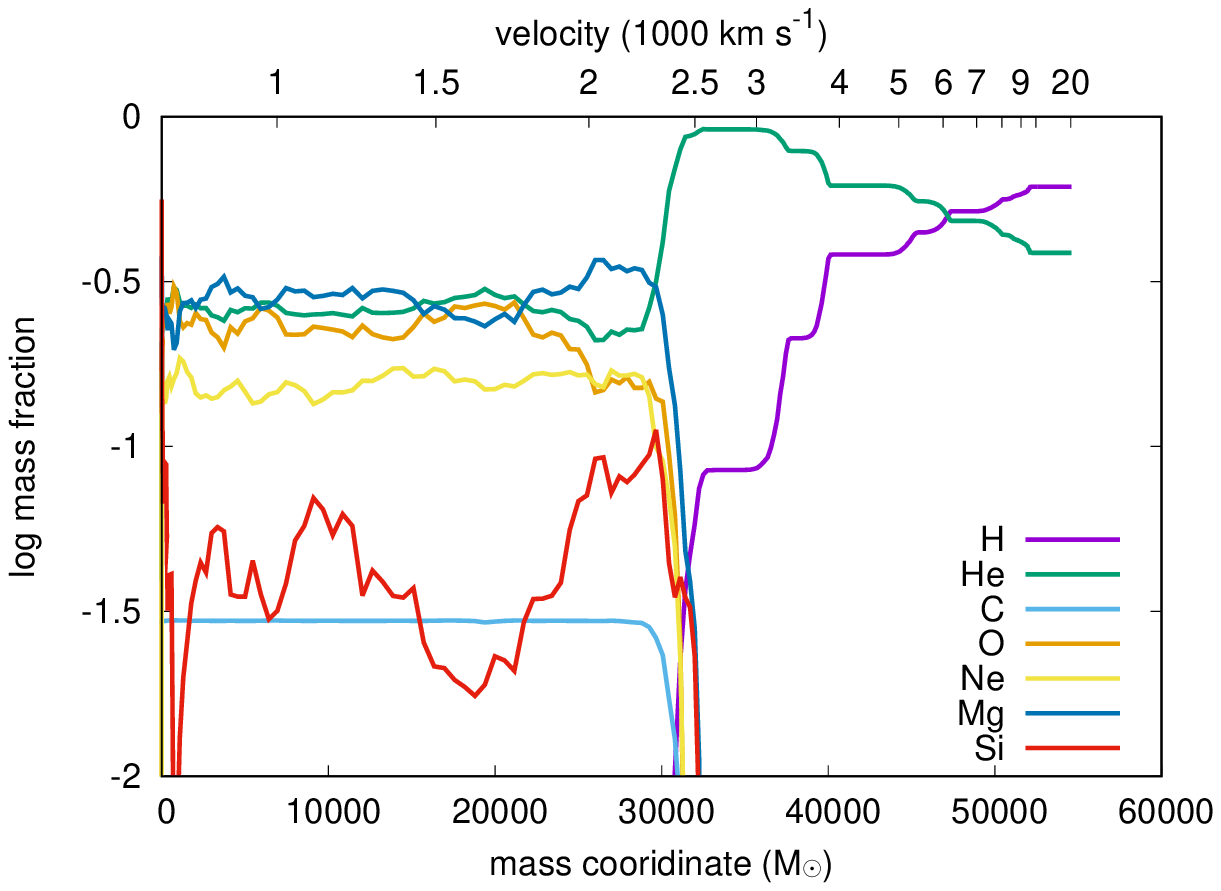}
    \caption{
    \textit{Top:}
    Initial density profile for the GRSN LC calculation by \texttt{STELLA}.
    \textit{Bottom:}
    Abundance profile for the GRSN LC calculation. The top axis shows the velocity of the mass coordinate during the homologous expansion.
    }
    \label{fig:abundance}
\end{figure}

We obtain the angle-averaged hydrodynamic and abundance profiles from the two-dimensional simulation and use it as the initial condition for our one-dimensional radiation hydrodynamics simulation described in the next section.
For each physical quantity at a given radius, we take ten directions from polar angles of $0.1\pi$, $0.2\pi$, $0.3\pi$, ..., $\pi$ in the two-dimensional result and then take the average value of the ten directions for the one-dimensional calculation. In this way we approximately take the effect of multi-dimensional mixing found in the two-dimensional calculation into account. Detailed chemical mixing patterns in the GRSN model are presented in Fig.~4 of \citet{chen2014gre}. 
The angle-averaged density and abundance profiles are shown in Fig.~\ref{fig:abundance}. A hydrogen+helium layer exists from the surface to around 31,000~\Msun. The layers below are mostly composed of helium, oxygen, neon, and magnesium.

\subsection{Light-curve calculation}
The numerical LC calculation is performed by using the one-dimensional radiation hydrodynamics code \texttt{STELLA} \citep{blinnikov1998sn1993j,blinnikov2000sn1987a,blinnikov2006sniadeflg}. In addition to the hydrodynamics equations, \texttt{STELLA} implicitly treats time-dependent equations of the angular moments of intensity averaged over a frequency bin with the variable Eddington method. \texttt{STELLA} calculates the spectral energy distributions (SEDs) at each time step and obtains multicolour LCs by convolving filter functions with the SEDs. We adopt the standard 100 frequency bins from 1~\AA\ to $5\times10^4$~\AA\ on a log scale. \texttt{STELLA} has been widely used for the LC modeling of hydrogen-rich SNe \citep[e.g.,][]{blinnikov2000sn1987a,baklanov2005,tominaga2011sb,moriya2011rsgwind,moriya2016rapidii,moriya2018iiplcs,goldberg2019,goldberg2020}, including hydrogen-rich Population~III SNe \citep{tolstov2016popiii,moriya2019popiii}. \texttt{STELLA} is also shown to provide very similar SN~II LCs to those obtained by the Monte Carlo radiation transfer approach \citep{tsang2020stellasedona}.

We put the initial condition that is shortly before the shock breakout provided by the \texttt{CASTRO} simulation and no energy is artificially injected. We do not set mass cut because no compact remnant remains in the GRSN explosion. 

Because we have the rest-frame SED information at each time step, we shift the SEDs at given redshifts and apply filter functions at the observer frame to estimate the observational properties of the GRSN at high redshifts. We adopt the following filter functions in NIR when we discuss the observational properties of the GRSN. The mean wavelength of each filter is also mentioned. The \textit{H} band filter ($1.77~\mathrm{\mu m}$) from Euclid, which is the reddest filter in Euclid, is adopted \citep{maciaszek2016euclid}. We take the \textit{J129} ($1.29~\mathrm{\mu m}$), \textit{H158} ($1.58~\mathrm{\mu m}$), and \textit{F184} ($1.84~\mathrm{\mu m}$) filters from the Nancy Grace Roman Space Telescope (RST, previously known as WFIRST)\footnote{\url{https://roman.gsfc.nasa.gov/science/Roman_Reference_Information.html}}. The \textit{K} band filter ($2.15~\mathrm{\mu m}$) is adopted from ULTIMATE-Subaru\footnote{\url{https://ultimate.naoj.org/english/}}, which is planned to have a similar \textit{K} band filter to the $K_S$ band filter of Subaru/MOIRCS\footnote{\url{https://subarutelescope.org/Observing/Instruments/MOIRCS/imag_sensitivity.html}}. We take three filters from Galaxy and Reionization EXplorer (G-REX), which is a proposed wide-field surveyor at $2-5~\mathrm{\mu m}$ (A.K. Inoue, private communication), i.e., the \textit{F232} ($2.32~\mathrm{\mu m}$), \textit{F303} ($3.03~\mathrm{\mu m}$), and \textit{F397} ($3.97~\mathrm{\mu m}$). These filters were previously proposed for the Wide-field Imaging Surveyor for High-redshift (WISH) satellite\footnote{\url{https://wishmission.org}} and are currently considered for G-REX. Finally, we take three broad filters from James Webb Space Telescope (JWST)/NIRCam\footnote{\url{https://jwst-docs.stsci.edu/near-infrared-camera/nircam-instrumentation/nircam-filters}}: \textit{F277W} ($2.77~\mathrm{\mu m}$), \textit{F356W} ($3.56~\mathrm{\mu m}$) and \textit{F444W} ($4.44~\mathrm{\mu m}$).

\begin{figure}
	\includegraphics[width=\columnwidth]{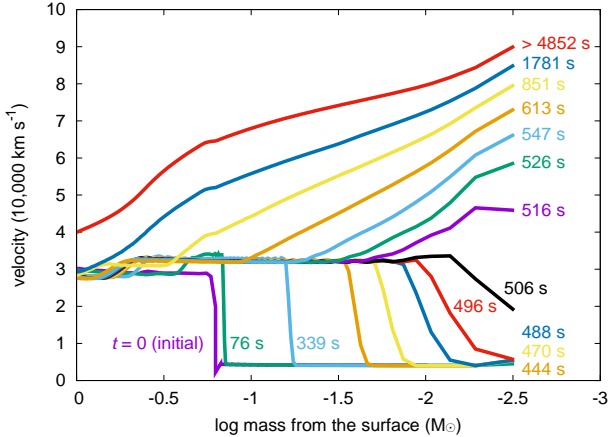}
    \caption{
    Velocity evolution at the surface of the progenitor shortly after the beginning of the \texttt{STELLA} calculation.
    }
    \label{fig:velociy}
\end{figure}

\begin{figure}
	\includegraphics[width=\columnwidth]{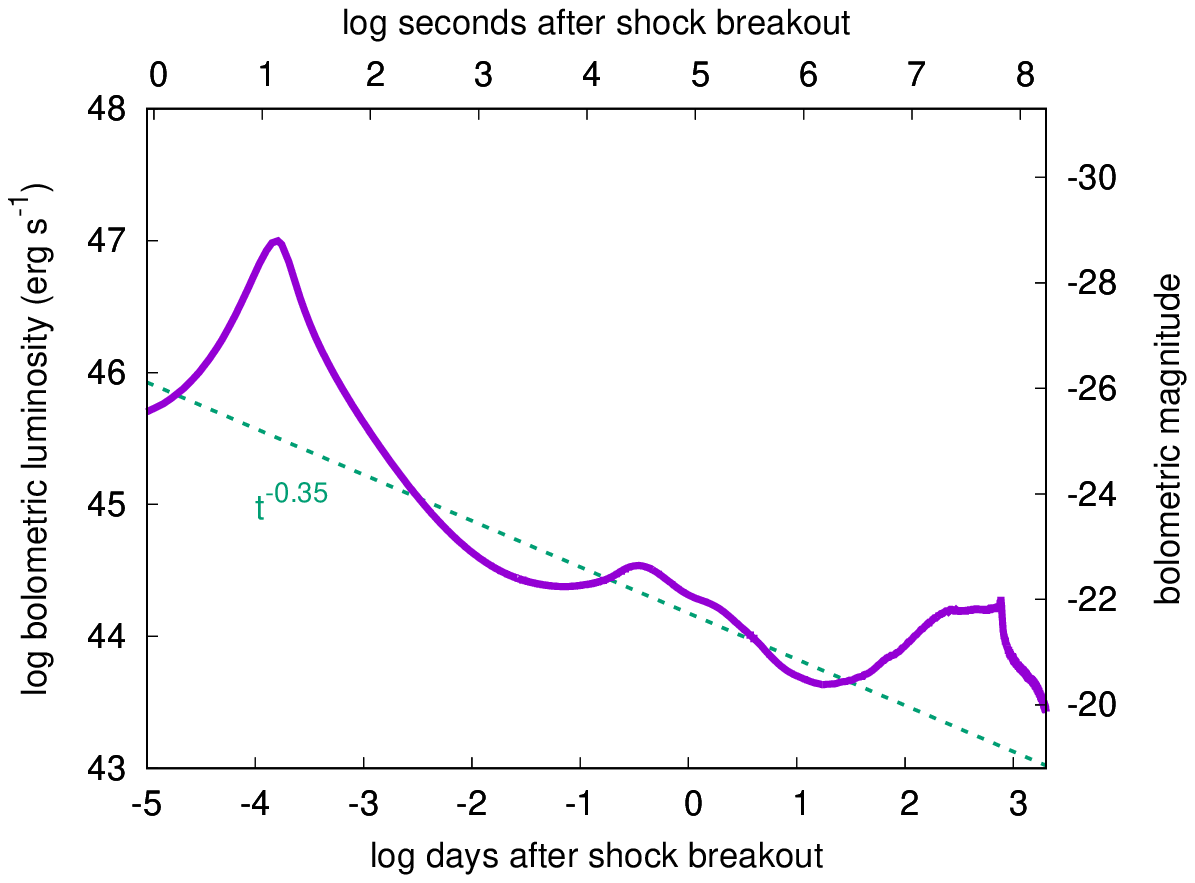}	
	\includegraphics[width=\columnwidth]{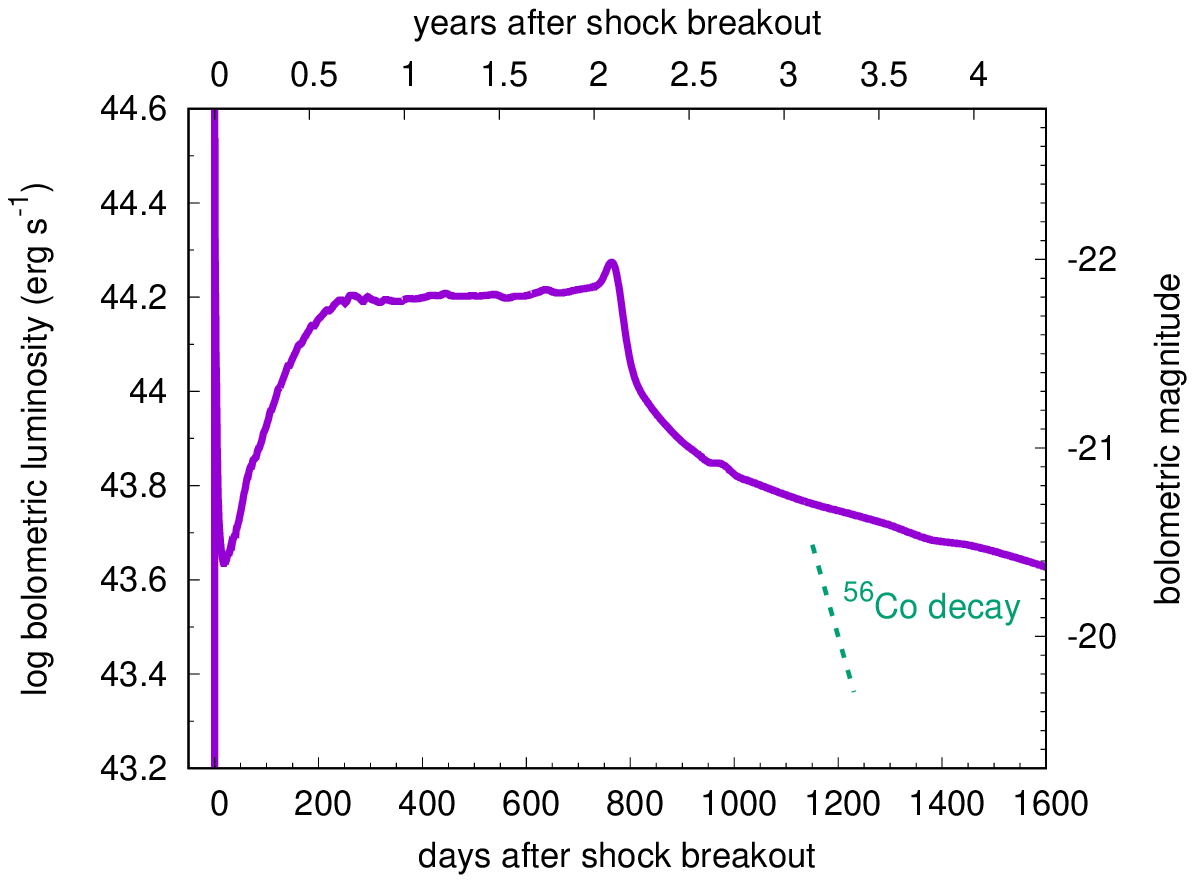}
    \caption{
    Bolometric LC of the GRSN. The horizontal axes in the top panel are in the log scale and those in the bottom panel are in the linear scale. The top panel shows the analytically estimated bolometric LC decay rate after the shock breakout ($\propto t^{-0.35}$) and the bottom panel shows the nuclear decay rate of \Co.
    }
    \label{fig:bolometric}
\end{figure}

\begin{figure}
	\includegraphics[width=\columnwidth]{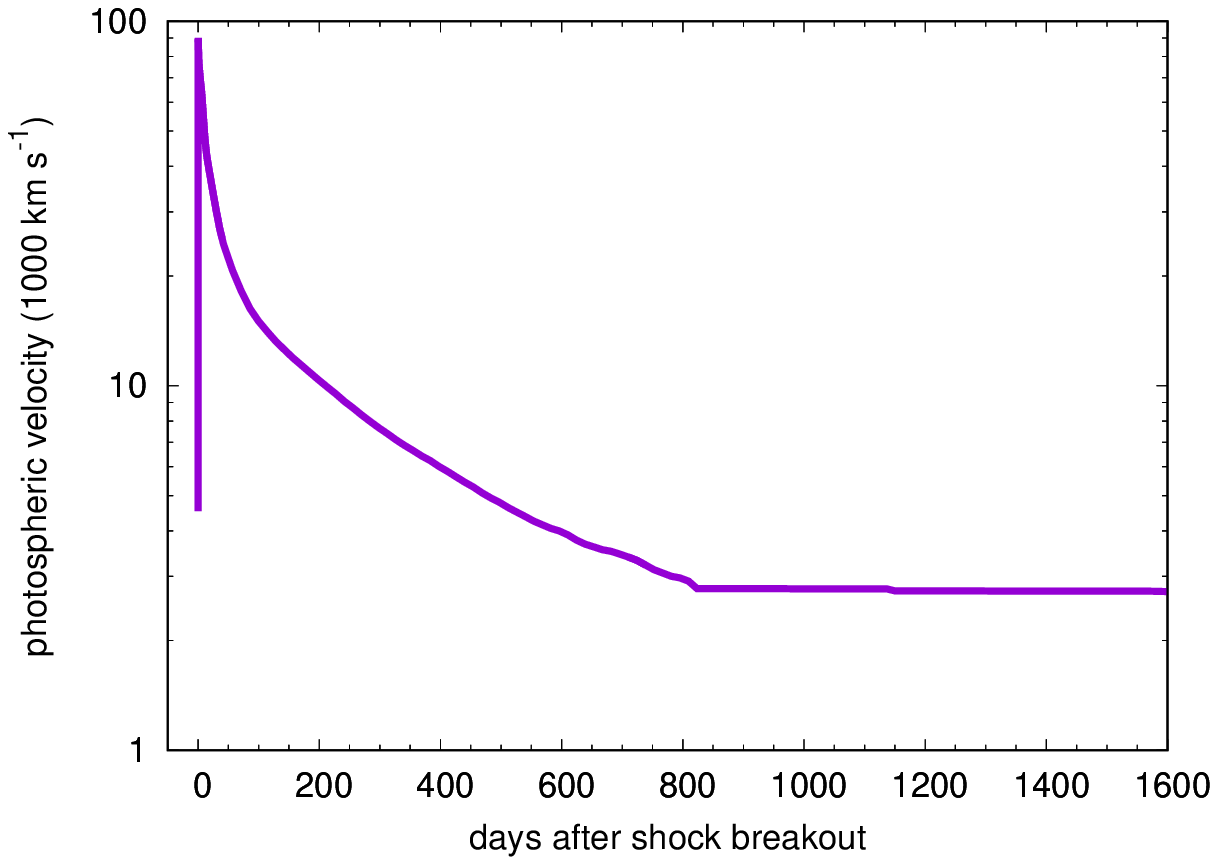}
	\includegraphics[width=\columnwidth]{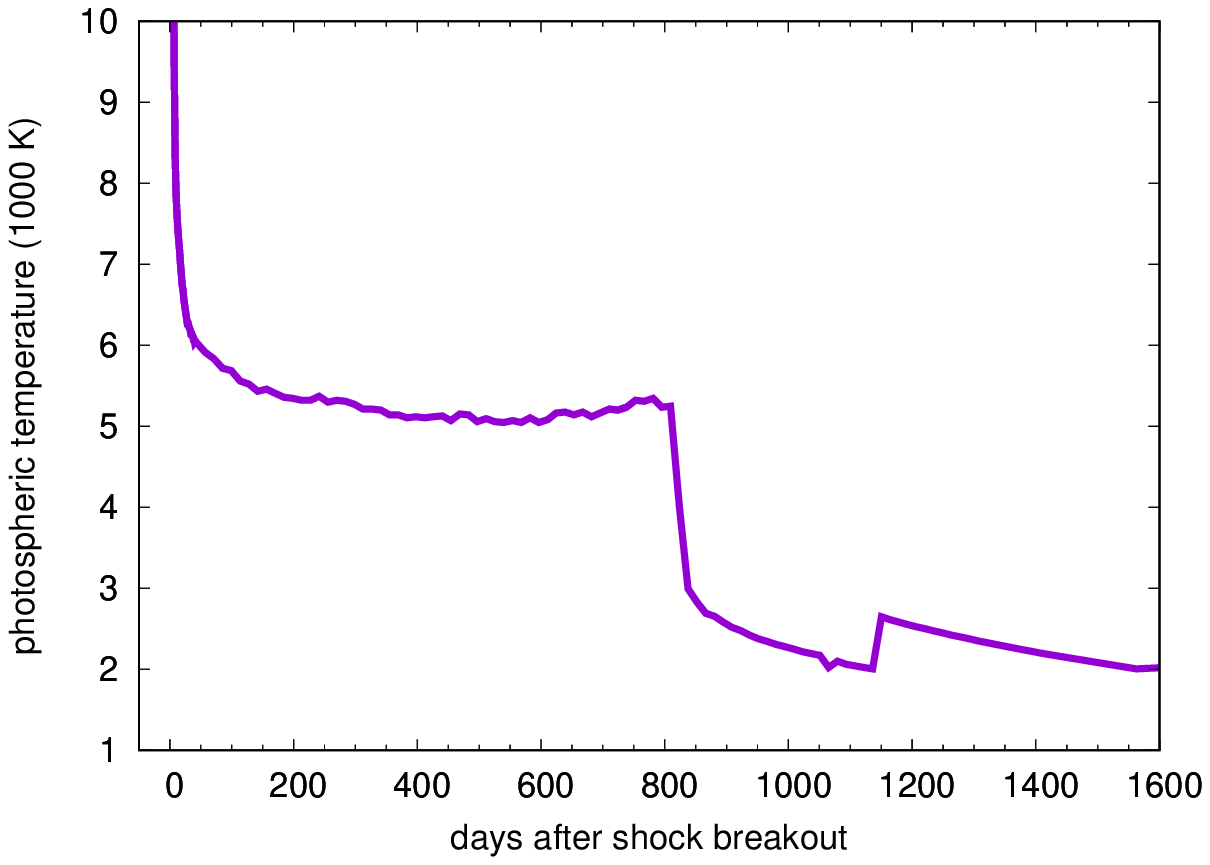}
	\includegraphics[width=\columnwidth]{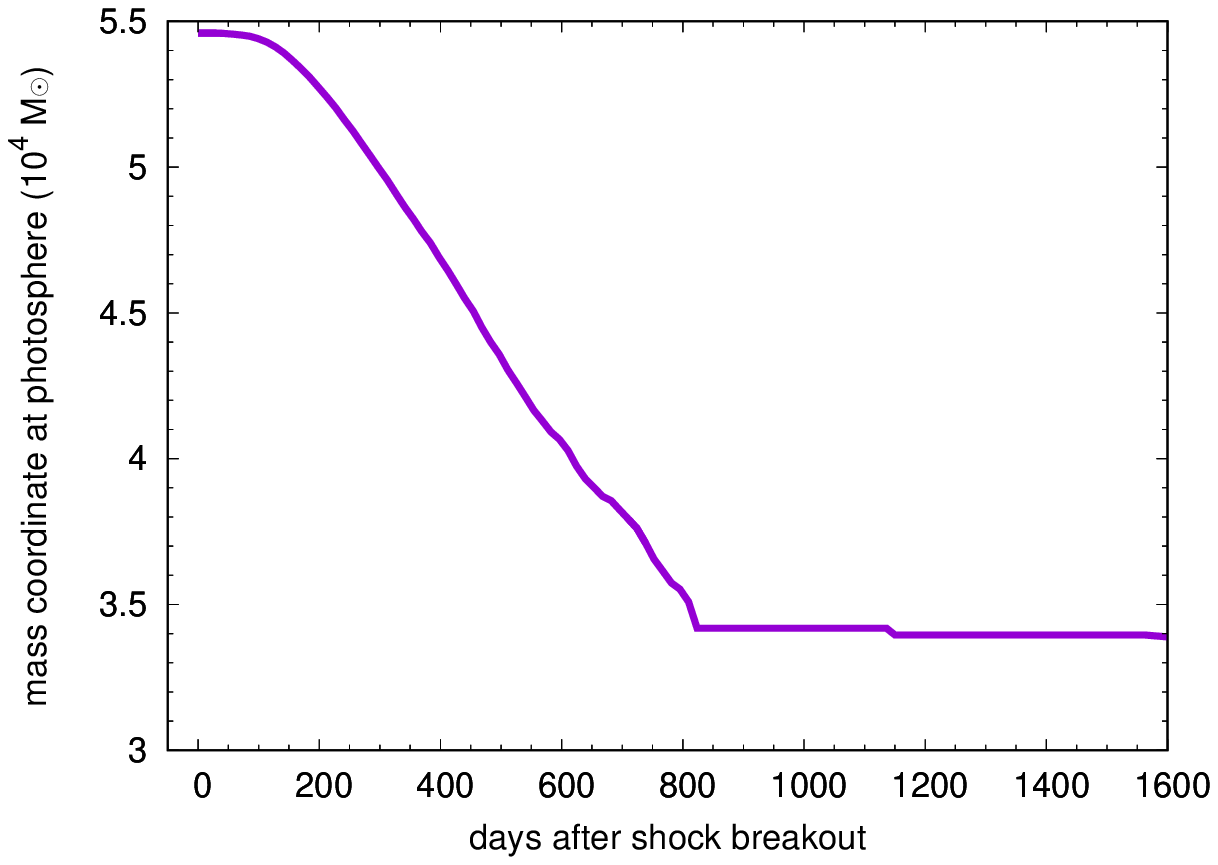}
	\includegraphics[width=\columnwidth]{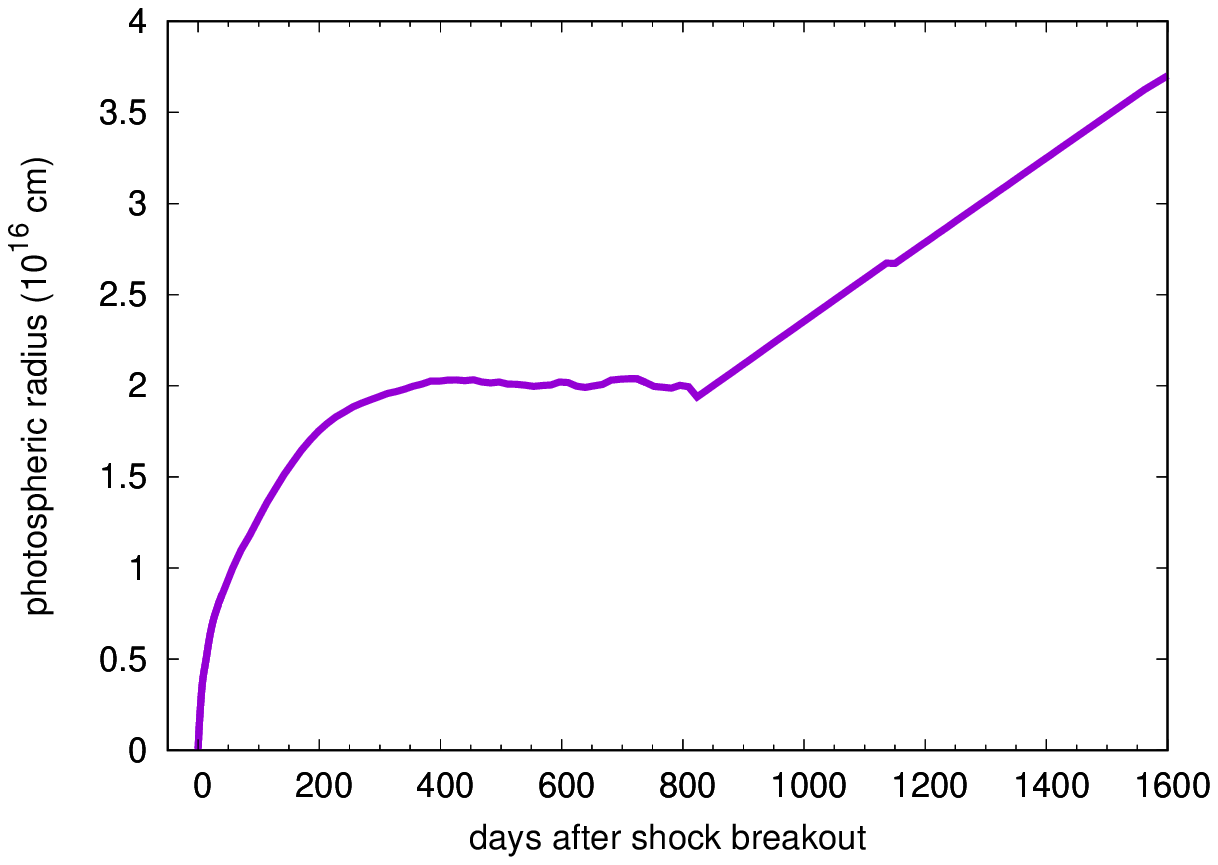}	
    \caption{
    Properties at the photosphere.
    }
    \label{fig:photosphere}
\end{figure}

\section{GRSN properties in the rest frame}\label{sec:restframeproperties}
Fig.~\ref{fig:velociy} shows the velocity evolution at the outermost layers at the beginning of the LC calculation. The initial condition ($t=0$) is shortly before the shock breakout and the shock front continues to travel towards the progenitor surface. The shock breakout occurs at $t\simeq 500~\mathrm{s}$. After the shock breakout, the homologous expansion of the ejecta is set at $t\simeq 5000~\mathrm{s}$ and the velocity of each mass shell is fixed from the moment. The velocity of the mass coordinate during the homologous expansion is shown in Fig.~\ref{fig:abundance}.

The bolometric LC of the GRSN is shown in Fig.~\ref{fig:bolometric}. Fig.~\ref{fig:photosphere} presents the velocity, temperature, mass coordinate, and radius of the photosphere which is where the Rosseland mean optical depth becomes $2/3$. We first see the shock breakout signal peaking at 14~s with $10^{47}~\mathrm{erg~s^{-1}}$. The shock breakout signal lasts for several hundred seconds which roughly correspond to the light-crossing time of the progenitor (600~s).

The bolometric LC evolution after the shock breakout roughly follows $\propto t^{-0.35}$ as expected from the analytical models  \citep[e.g.,][]{chevalier2008,rabinak2011,kozyreva2020}.
The bolometric LC keeps declining until 17~d when the outermost layers get close to the hydrogen recombination temperature at $\simeq 6000-5000~\mathrm{K}$. The subsequent bolometric LC evolution can be interpreted simply with $L\propto R^2T^4$, where $L$ is the bolometric luminosity, $R$ is the photospheric radius, and $T$ is the photospheric temperature. The bolometric luminosity increases from 17~d to 250~d when the recession velocity of the photosphere is slower than the hydrodynamic expansion velocity of the ejecta. In other words, the photospheric temperature is set at the recombination temperature but the photospheric radius increases (Fig.~\ref{fig:photosphere}), making the bolometric luminosity to increase.

The bolometric luminosity increases until 250~d when the recession velocity of the photosphere and the hydrodynamic expansion velocity matches. From this time, the photosphere is kept at the same radius and it also keeps the same recombination temperature. Thus, the bolometric luminosity becomes constant and the plateau phase appears. This physical condition is the same as that found in Type~IIP SNe during their plateau phase \citep{grassberg1971,falk1977}. The plateau luminosity ($\simeq 1.5\times 10^{44}~\mathrm{erg~s^{-1}}$) is much larger than those observed in Type~IIP SNe \citep{bersten2009} because of the extremely larger explosion energy. Exceeding $10^{44}~\mathrm{erg~s^{-1}}$, the GRSN can be regarded as a member of superluminous SNe \citep[][for recent reviews]{moriya2018slsnrev,gal-yam2019slsnrev}. The plateau continues up to around 800~d when the photosphere reaches the bottom of the layers containing hydrogen (see the mass coordinate of the photosphere in Fig.~\ref{fig:photosphere}). The hydrogen recombination ends at this point and the photospheric temperature drops suddenly. The plateau duration is about 550~d in the rest frame which is also much longer than those found in Type~IIP SNe (about 100~d, \citealt{bersten2009,anderson2014}). The total radiated energy during the plateau phase amounts to $\simeq 7\times 10^{51}~\mathrm{erg}$.

The plateau phase is followed by the sudden drop in luminosity and then the ``tail'' phase starts. The bolometric luminosity keeps decreasing during the tail phase. The tail phase luminosity is not powered by the radioactive decay of \Co\ unlike the case of Type~IIP SNe because of the small amount of \Ni\ (less than 0.1~\Msun) in the ejecta. 
The LC decline rate is different from that of \Co\ as shown in Fig.~\ref{fig:bolometric}.
The luminosity source in this tail phase is still thermal energy from the initial explosion as in the previous hydrogen recombination phase. About a half of magnesium, which is the most abundant element in the hydrogen-free core (Fig.~\ref{fig:abundance}), remain ionized even after the plateau phase because of its low ionization energy. Therefore, the opacity at the hydrogen-free core remains high (of the order of $0.01~\mathrm{cm^2~g^{-1}}$) even after the hydrogen recombination phase and the photosphere is kept at the surface of the hydrogen-free layers during the epochs presented in Fig.~\ref{fig:bolometric}. This is why the photospheric radius increase linearly after the hydrogen recombination phase, i.e., the photosphere is kept at the surface of the hydrogen-free core. The bolometric luminosity decreases as the temperature decreases due to the adiabatic cooling. In the photospheric temperature evolution, we find a sharp increase at around 1150~days. The exact reason for the jump is not clear and it could be a numerical artifact.

Our bolometric LC behavior is different from that presented in the previous study in \citet{whalen2013grsn} in which the same initial explosion model is used but the numerical LC calculation is performed in a different code. The bolometric LC from \citet{whalen2013grsn} shows a slow luminosity decline starting from the shock breakout peak. Their LC model does not show the plateau phase caused by the hydrogen recombination as we found in our model. We believe that the existence of the plateau phase is expected from the presence of the massive hydrogen-rich envelope as in the case of Type~IIP SNe \citep{grassberg1971,falk1977}. We have compared our results with predictions of the analytic model by \citet{nagy2016} (see also \citealt{szalai2019} for the update of that model). We find a good agreement of synthetic light curves produced by \texttt{STELLA} with this simple model on the plateau stage.
The earlier bolometric LC behaivor in our model is also consistent with the analytic models as discussed earlier in this section. The reasons for the difference between our LC model and the model in \citet{whalen2013grsn} remain unclear.

\begin{figure}
	\includegraphics[width=\columnwidth]{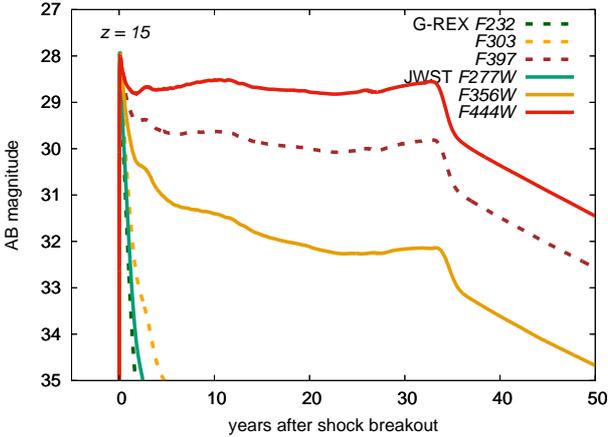}	
    \caption{
    GRSN LCs at $z=15$ with the G-REX and JWST filters in the observer frame.
    }
    \label{fig:highredshiftlc}
\end{figure}

\section{Discovering GRSNe in the future NIR surveys}\label{sec:howtodiscover}
We discuss the observational properties of GRSNe when they appear in the early Universe based on the GRSN model presented in the previous section. 

\subsection{GRSNe as persistent sources}
\begin{figure}
	\includegraphics[width=\columnwidth]{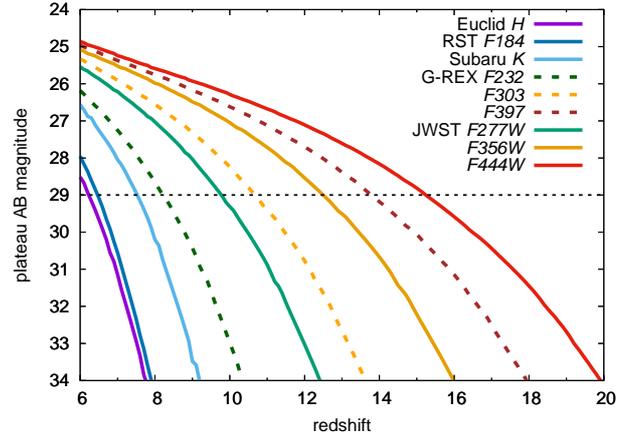}
    \caption{
    Magnitudes of the GRSN at the plateau phase as a function of redshift. The SED at 500~d after shock breakout presented in Section~\ref{sec:restframeproperties} is used to estimate the plateau magnitudes.
    The horizontal dashed line shows a typical limiting magnitude of planned imaging surveys (29~mag).
    }
    \label{fig:redshiftmagnitude}
\end{figure}

Fig.~\ref{fig:highredshiftlc} shows the GRSN LCs at $z=15$ with the G-REX and JWST filters.
The observed wavelengths correspond to $0.1-0.3~\mathrm{\mu m}$ in the rest frame. The SEDs at $0.1-0.3~\mathrm{\mu m}$ are strongly affected by metal absorption and have steep gradients. This causes large differences in the observed magnitudes even if the differences in the filter central wavelengths are small. This effect of absorption becomes larger with higher redshifts (cf. Fig.~\ref{fig:redshiftmagnitude}).

The most prominent feature is the long plateau phase lasting for three decades. It originate from the luminous plateau phase lasting for 550~d in the rest frame (Section~\ref{sec:restframeproperties}). Because of the time dilution, the long plateau phase becomes even longer and lasts for decades when GRSNe appear at high redshifts.

The long plateau lasting for more than a decade would be observed as a persistent source rather than a transient during the NIR survey, because the major NIR survey missions are conducted by satellite telescopes and their operation period is usually set to 5~years. Fig.~\ref{fig:redshiftmagnitude} shows the predicted magnitudes of the GRSN plateau as a function of redshift. We find that GRSNe up to $z\simeq 15$ can be observed with G-REX and JWST if a deep imaging survey reaching at least $\simeq 29~\mathrm{AB~mag}$ is conducted. The deeper images enable us to reach higher redshifts (Fig.~\ref{fig:redshiftmagnitude}).

\begin{figure}
	\includegraphics[width=\columnwidth]{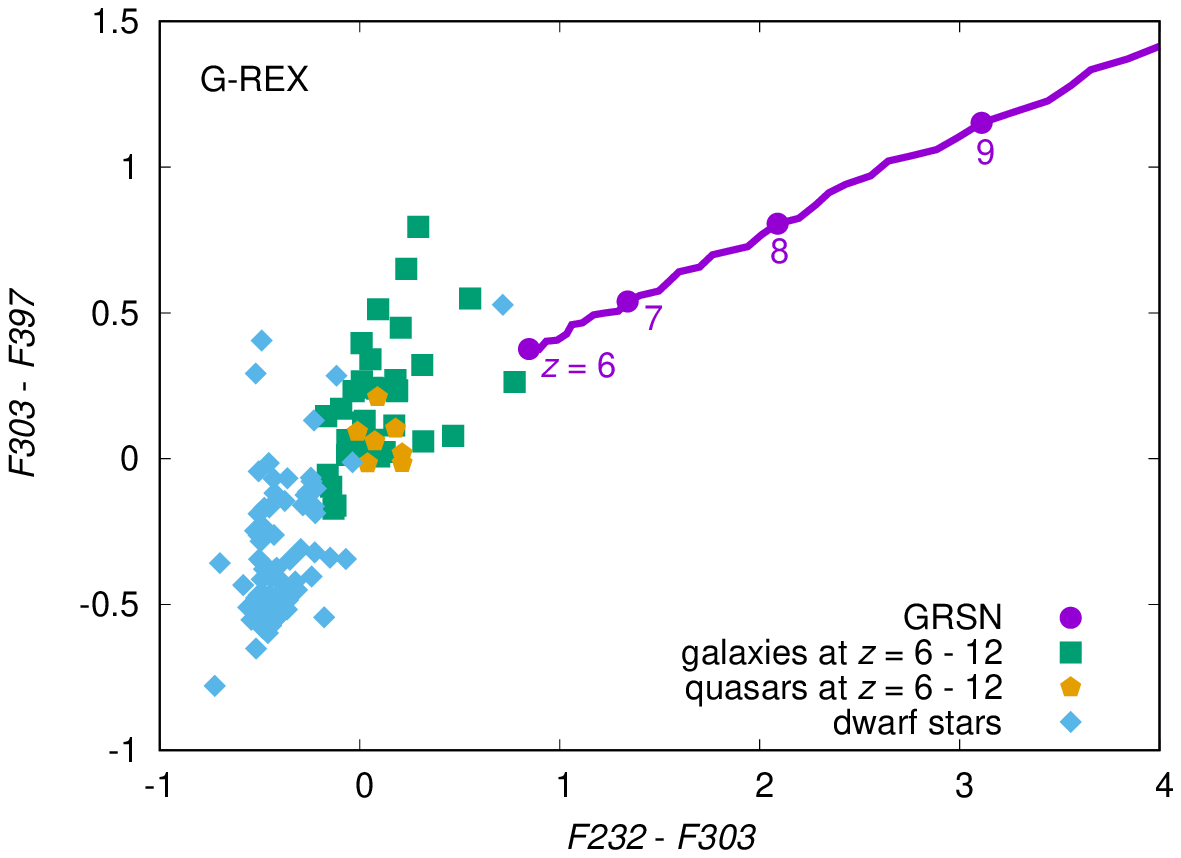}	
	\includegraphics[width=\columnwidth]{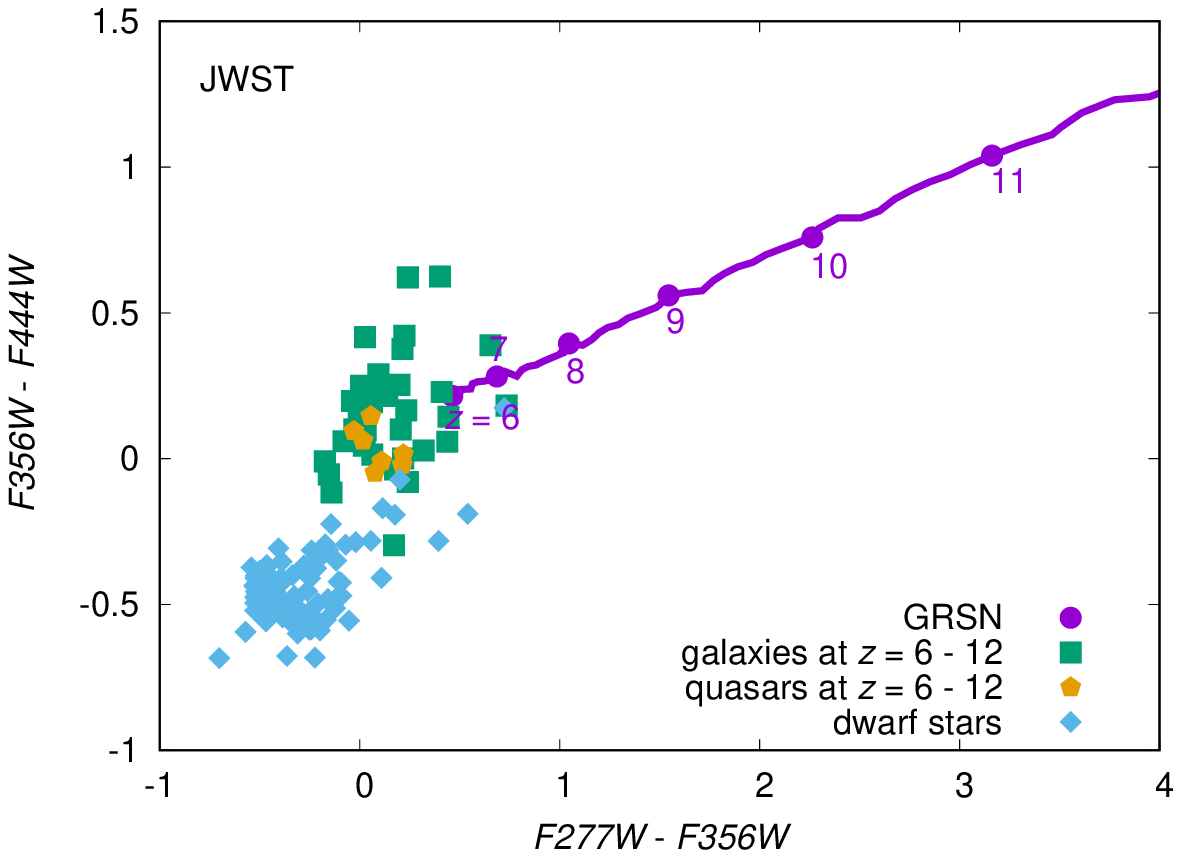}		
    \caption{
    Color-color diagrams of the GRSN at the plateau phase (500~d after the shock breakout). We also put high-redshiift galaxies and dwarf stars in our Galaxy for comparison.
    }
    \label{fig:color}
\end{figure}

To identify non-transient GRSNe, we need to distinguish them from other persistent sources such as high-redshift galaxies. Fig.~\ref{fig:color} shows the color-color diagrams in which GRSNe at plateau, high-redshift galaxies, high-redshift quasars, and dwarf stars in our Galaxy. 
The high-redshift galaxy templates are produced
with the stellar radiation provided by BPASS (v2.1; \citealt{eldridge2017}) in addition to nebular emission (both lines and continua) as calculated in a self consistent manner with Cloudy (see \citealt{nakajima2018} for more details). Here we adopt ``young'' and ``evolved'' galaxies, with a continuous star formation history at the ages of $10$ and $500$~Myr, the metallicity of $0.1$ and $0.5$ solar metallicity, and the dust reddening of $E(B-V)=0.0$ and $0.2$ assuming the \citet{calzetti2000}'s extinction law, respectively.
The young population shows intense emission lines
with a rest-frame equivalent width of [OIII]$\lambda 5007$+H$\beta$
as large as $\sim1000$\,\AA, which corresponds to extreme 
emission line galaxies such as green pea galaxies 
\citep[e.g.,][]{cardamone2009green}.
The evolved population represents continuum-selected 
galaxies like Lyman-break galaxies \citep[e.g.,][]{steidel2014}.
Both galaxy populations are then placed at redshifts of $z=6-12$ with an IGM attenuation prescribed by \citet{inoue2014}. At these redshifts, strong optical emission lines such as H$\alpha$ and [O\,{\sc iii}] $\lambda 5007$ fall in the NIR filters of G-REX and JWST/NIRCam. The nebular component in our galaxy-SEDs is thus essential to obtain realistic predictions of NIR colors of high-redshift galaxies, which are inferred to present intense H$\alpha$ and [O\,{\sc iii}]$+$H$\beta$ emission lines (e.g., \citealt{smit2014,roberts-borsani2016}).
The quasar models are based on a composite of spectra of quasars at $z=1-2$ observed in the rest-frame ultraviolet to near-infrared wavelength \citep{selsing2016}. 
The composite is redshifted and IGM-attenuated in the same way as the galaxy models to mock up the high-redshift quasars.
The dwarf stars are plotted by using the NIR spectra obtained by SpeX \citep{rayner2003specxinst} and presented in \citep{cushing2005specx,rayner2009specx}\footnote{We obtained the spectra at \url{http://irtfweb.ifa.hawaii.edu/~spex/IRTF_Spectral_Library/}.}. We took all the dwarf spectra with the spectral coverage up to $4~\mathrm{\mu m}$ for G-REX and $5~\mathrm{\mu m}$ for JWST and convolved the filter functions. When there is a gap in the dwarf spectra caused by the telluric absorption, we simply interpolated the gap.

The color-color diagram (Fig.~\ref{fig:color}) shows that GRSNe are much redder than high-redshift galaxies, high-redshift quasars, and dwarf stars and they can be easily distinguished from GRSNe at the plateau phase. This is because the temperature of the plateau is only $\simeq 5000-6000~\mathrm{K}$ in the rest frame and they become extremely red if they are at high redshifts. This temperature at the plateau phase is determined by the hydrogen recombination temperature. Thus, the color during the plateau phase is a solid prediction that does not depend much on the actual mass and energy of the GRSN explosions.

\begin{figure}
	\includegraphics[width=\columnwidth]{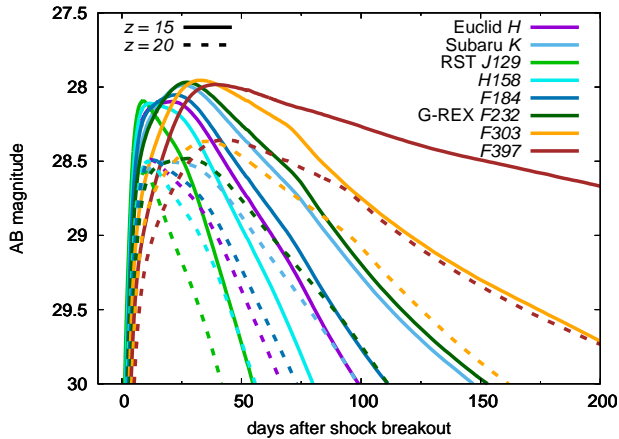}
    \caption{GRSN LCs at $z=15$ and 20 during the cooling phase after the shock breakout with the selected NIR filters.}
    \label{fig:lc_at_cooling}
\end{figure}

\begin{figure}
	\includegraphics[width=\columnwidth]{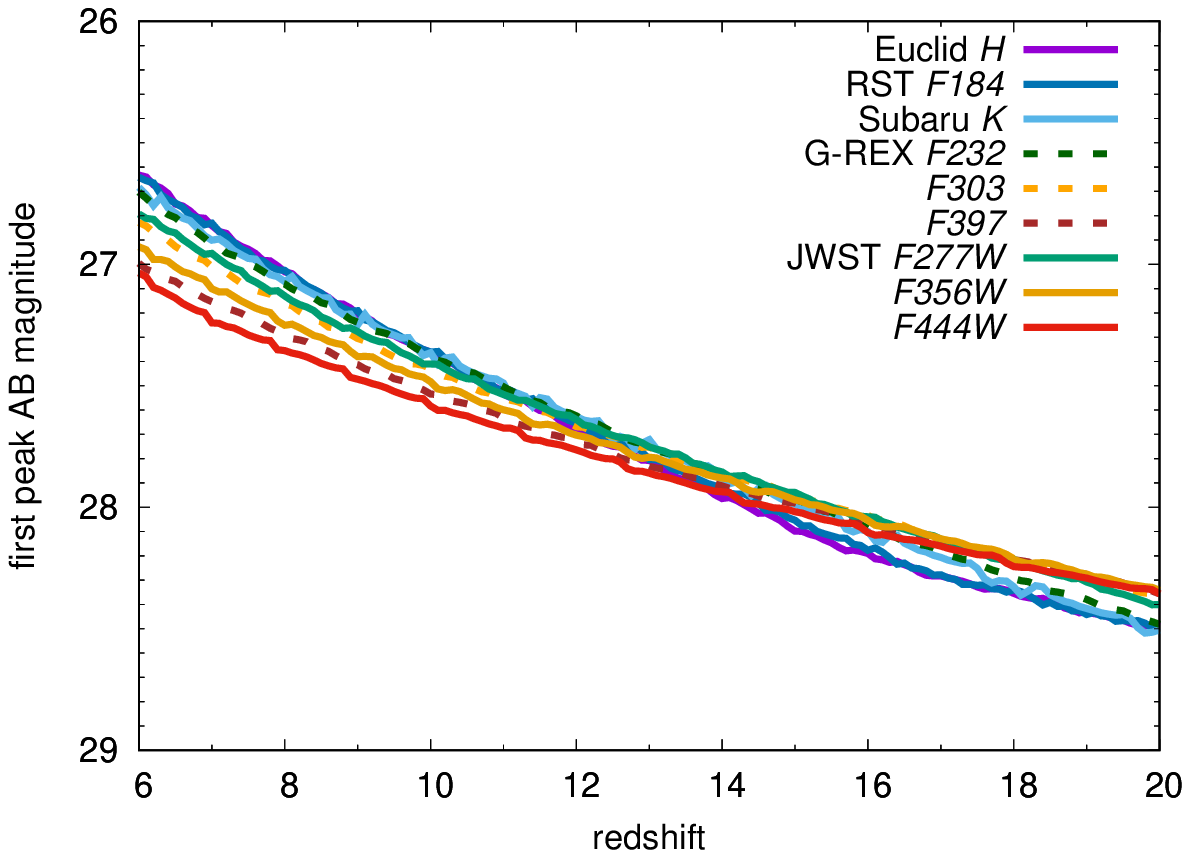}
	\includegraphics[width=\columnwidth]{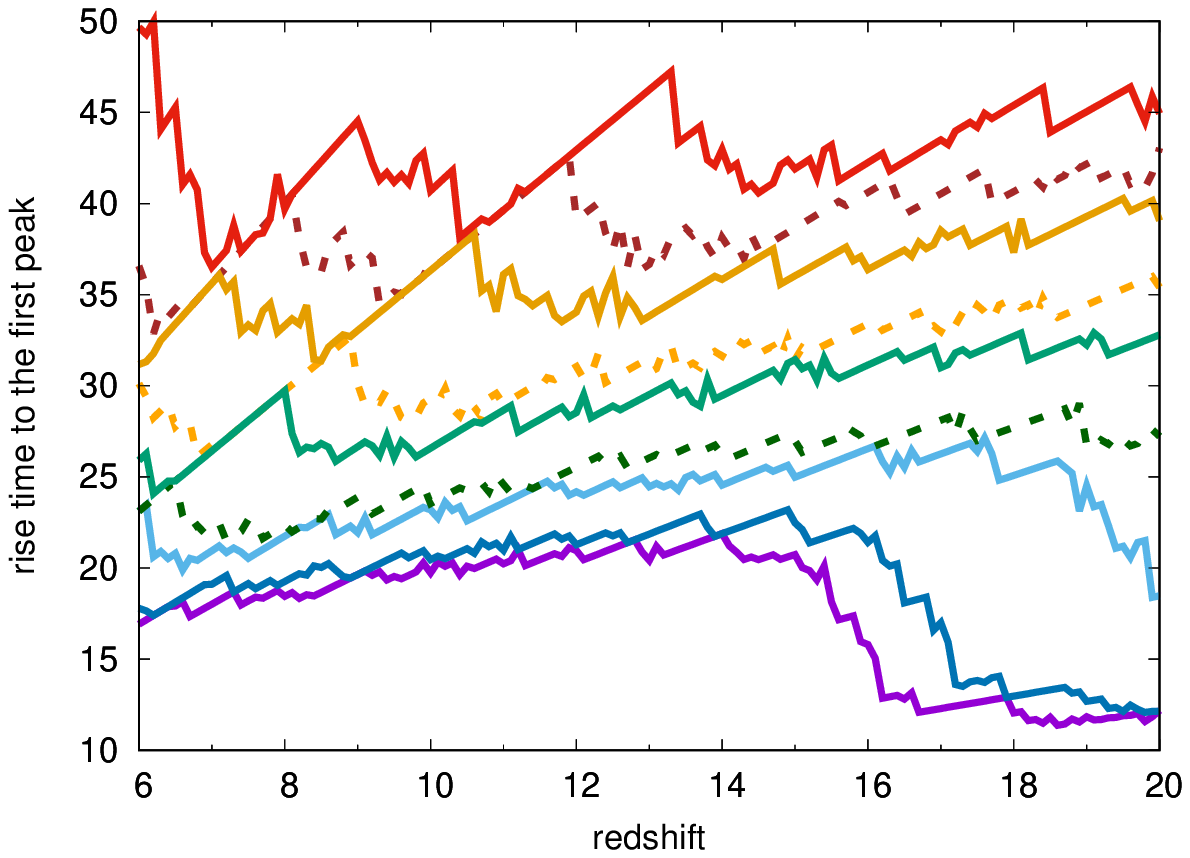}	
    \caption{
    Peak magnitudes and rise times of GRSNe during the cooling phase shortly after the shock breakout as a function of redshift.
    }
    \label{fig:redshiftfirstpeak}
\end{figure}

\subsection{Discovering GRSNe with NIR transient surveys}
As discussed previously by \citet{whalen2013grsn}, it is also possible to find GRSNe with NIR transient surveys. As presented in Fig.~\ref{fig:bolometric}, GRSNe are brightest shortly after the shock breakout and their luminosity keeps declining until 17~d. GRSNe at this phase can be observed at $z\gtrsim 15$ thanks to their large luminosity and high temperature. Fig.~\ref{fig:lc_at_cooling} shows examples of the GRSN LCs during the declining phase. The LCs rise in about a month and declines. Fig.~\ref{fig:redshiftfirstpeak} shows the peak luminosity of the first peak and the rise time to the peak in the NIR filters as a function of redshift.  NIR transient surveys reaching 28.5 AB~mag are required to find high-redshift GRSNe. The timescale is about a month. Thus, deep NIR transient surveys reaching down to 28.5 AB~mag with a cadence of around 10~d are required to identify high-redshift GRSNe shortly after the explosion.

\section{Conclusions}\label{sec:summary}
We have presented the observational properties of the GRSN from the 55,500~\Msun\ Population~III star. The progenitor has the radius of 256~\Rsun\ and the explosion energy is $9\times 10^{54}~\mathrm{erg}$. We take the results of the GRSN explosion simulation up to shortly before the shock breakout conducted previously by \citet{chen2014gre}. We put it as the initial condition to the radiation hydrodynamics code \texttt{STELLA} to investigate its observational properties. The overall observational properties of the GRSN is similar to those of Type~IIP SNe, but the GRSN has the much longer (550~d in the rest frame) and more luminous ($1.5\times 10^{44}~\mathrm{erg~s^{-1}}$) plateau phase. The photospheric temperature during the plateau is characterized by the hydrogen recombination temperature ($\simeq 5000-6000~\mathrm{K}$) as is the case for Type~IIP SNe.
The plateau phase was not predicted to exist in the previous study by \citet{whalen2013grsn} in which the same initial explosion model is used to investigate the GRSN properties, but the existence of the plateau phase is expected from the presence of the massive hydrogen-rich envelope as in the case of Type~IIP SNe. The plateau phase lasts for many decades when the GRSN appears at high redshifts and the GRSN is not recognized as a transient in the future NIR imaging surveys. However, it can be distinguished from other persistent sources such as high-redshift galaxies and local dwarf stars by using color information. The deep NIR images reaching 29 AB~mag obtained by G-REX and JWST will allow us to search for the GRSN up to $z\simeq 15$. The deeper images allow us to reach higher redshifts. The NIR transient surveys reaching $28-28.5~\mathrm{AB~mag}$ with a cadence of 10~d may identify the GRSN at $z\gtrsim 15$ during the adiabatic cooling phase after the shock breakout as a transient.

Because the exact mass range of the GRSN explosions is unclear, it is difficult to predict the expected number of the GRSN detection in the deep NIR images obtained by the future NIR surveys. Still, the extremely red color of high-redshift GRSNe makes them easy to identify. The existence or lack of the GRSNe can be constrained in the future deep NIR images without conducting transient surveys. Such a constraint provides valuable information on the fate of SMSs possibly leading to the formation of SMBHs in the early Universe.

\section*{Acknowledgements}
We thank the anonymous referee for constructive comments that improved this paper.
T.J.M. thanks the organisers of the First Star VI conference at Universidad de Concepci\'on, Chile, where this work was initiated.
T.J.M. is supported by the Grants-in-Aid for Scientific Research of the Japan Society for the Promotion of Science (JP18K13585, JP20H00174).
K.C. acknowledges support from the Ministry of Science and Technology (Taiwan, R.O.C.) grant number MOST 107-2112-M-001-044-MY3.
S.B. is supported  by grant RSF 19-12-00229 for the development of
STELLA code.
This research has been supported in part by the RFBR (19-52-50014)-JSPS bilateral program.
Numerical computations were in part carried out on PC cluster at Center for Computational Astrophysics (CfCA), National Astronomical Observatory of Japan.

\section*{Data availability}
The data underlying this article will be shared on reasonable request to the corresponding author.

%%%%%%%%%%%%%%%%%%%%%%%%%%%%%%%%%%%%%%%%%%%%%%%%%%



\bibliographystyle{mnras}
\bibliography{mnras} % if your bibtex file is called example.bib

%\begin{thebibliography}{99}
%\end{thebibliography}

%%%%%%%%%%%%%%%%%%%%%%%%%%%%%%%%%%%%%%%%%%%%%%%%%%

%%%%%%%%%%%%%%%%% APPENDICES %%%%%%%%%%%%%%%%%%%%%

%\appendix


%%%%%%%%%%%%%%%%%%%%%%%%%%%%%%%%%%%%%%%%%%%%%%%%%%

% Don't change these lines
\bsp	% typesetting comment
\label{lastpage}
\end{document}